\documentclass[11pt,twoside
]{article}

\usepackage{baaa2009}
\usepackage{graphicx}
\usepackage{subfigure}
\usepackage{psfrag}
\usepackage{amssymb}
\usepackage[spanish,activeacute,english]{babel}
\usepackage[latin1]{inputenc}
\usepackage[T1]{fontenc} % Computer Modern (CM) fonts
\usepackage{ae,aecompl} % and: dvips -Pcmz -Pamz macros_aaa.dvi
\usepackage{latexsym}
\usepackage{verbatim}
\usepackage{amsmath}
\usepackage{stmaryrd}
\usepackage{amsfonts}
\usepackage{amssymb}
\usepackage{wasysym}
\usepackage[colorlinks=true,dvips]{hyperref}
\begin{document}
%%%%%%%%%%%%%%%%%%%%%%%%%%%%%%%%%%%%%%%%%%%%%%%%%%%%%%%%%%%%%%%%%%%%%%%%%%
%%%% SELECCIONE EL IDIOMA EN QUE SE ESCRIBE EL ART?CULO:              %%%%
%\myselectspanish
\myselectenglish
%%%%%%%%%%%%%%%%%%%%%%%%%%%%%%%%%%%%%%%%%%%%%%%%%%%%%%%%%%%%%%%%%%%%%%%%%%
\vskip 1.0cm
\markboth{Noelia Jim\'enez et al.}%
{Brigth end of the CMR}

\pagestyle{myheadings}
%%%% DESCOMENTE LA LINEA QUE DESCRIBE EL CARACTER DE SU TRABAJO       %%%%
\vspace*{0.5cm}
%\noindent TRABAJO INVITADO 
\noindent PRESENTACI\'ON ORAL
%\noindent PRESENTACI\'ON MURAL
%\noindent RESUMEN 
\vskip 0.3cm
\title{The Bright End of the Colour-Magnitude Relation}

%\title{ Template paper for publication in the Bulletin of the 
%Argentinian Astronomical Association with instructions for the use of 
%\LaTeX{}}

\author{ Noelia Jim\'enez$^{1,2,3}$, Sof\'ia A. Cora$^{1,2,3}$, Lilia P. Bassino$^{1,2,3}$ \& Anal\'ia Smith Castelli$^{1,2,3}$}

\affil{ (1) Facultad de Ciencias Astron\'omicas y
  Geof\'isicas(FCAG-UNLP)\\ (2) Consejo Nacional de Investigaciones
  Cient\'ificas y T\'ecnicas(CONICET)\\ (3) Instituto de Astrof\'isica
  de La Plata (CCT La Plata,CONICET)\\ }

\begin{abstract}
 We investigate the origin of the colour-magnitude relation (CMR)
followed by early-type cluster galaxies by using a combination of
cosmological {\em N}-body simulations of cluster of galaxies and a
semi-analytic model of galaxy formation (Lagos, Cora \& Padilla
2008). Results show good agreement between the general trend of the
simulated and observed CMR.  However, in many clusters, the most
luminous galaxies depart from the linear fit to observed data
displaying almost constant colours.  With the aim of understanding
this behaviour, we analyze the dependence with redshift of the stellar
mass contributed to each galaxy by different processes, i.e.,
quiescent star formation, and starburst during major/minor and wet/dry
mergers, and disk instability events. The evolution of the
metallicity of the stellar component, contributed by each of these
processes, is also investigated.  We find that the major contribution
of stellar mass at low redshift is due to minor dry merger events,
being the metallicity of the stellar mass accreted during this process
quite low.  Thus, minor dry merger events seem to increase the mass of
the more luminous galaxies without changing their colours.

\end{abstract}

\begin{resumen}
Investigamos el origen de la relaci\'on color magnitud (RCM) observada
en galaxias de tipo temprano residentes en c\'umulos, 
combinando una simulaci\'on cosmol\'ogica de
{\em N}-cuerpos y un modelo semianal\'itico de formaci\'on de galaxias
(Lagos, Cora \& Padilla 2008).  Obtenemos un buen acuerdo entre la
tendencia de la RCM simulada y observada.  Sin embargo, en mucho c\'umulos,
las galaxia
m\'as luminosas de la relaci\'on se separan del ajuste lineal a los datos
observados, mostrando colores casi constantes. Con el objetivo de
entender este comportamiento, analizamos la dependencia con el
corrimiento al rojo de la masa estelar aportada a cada galaxia por
diferentes procesos: formaci\'on de estrellas debido al gas fr\'io
disponible, y brotes estelares durante fusiones menores/mayores y
secas/h\'umedas, y durante eventos
de inestabilidad de disco. Se investig\'o tambi\'en la evoluci\'on de la
metalicidad de la componente de masa estelar contribuida por cada uno
de estos procesos. Encontramos que la mayor contribuci\'on a la masa
estelar de las galaxias es debida a las fusiones secas menores, siendo
bastante baja la metalicidad de la masa estelar acretada por la
galaxia. De este modo, las fusiones secas menores parecen agregar masa
estelar a las galaxias m\'as masivas sin alterar los colores de las mismas.

\end{resumen}

\section{Introduction}

It has long been known that there exists a bimodal distribution of
galaxies in the colour-magnitude plane, separated into a tight
colour-magnitude relation (CMR) and a ``blue cloud''.  The first one,
also known as the ``red sequence'', is populated prototypically by
early-type galaxies which are gas-poor and have low levels of star
formation, while late-type galaxies are typical objects of the blue
cloud. The colour-magitude relation can be understood as a
mass-metallicity relation; the more luminous, and consequently, the
more massive galaxies in this relation have deep potencial wells
capable of retaining the metal content released by supernovae events
and stellar winds.  The CMR seems to be quite universal since it is
followed by galaxies in the field as well as in groups and clusters
(Bower et al. 1992), being the fraction of red galaxies larger in
denser environments.  Generally, a linear relation has been used to
fit the correlation between luminosity and colour of cluster galaxies
lying in the red sequence; however, different fits have been suggested
(e.g., Janz \& Lisker 2009).  The CMR constitutes one of the major
tools for testing galaxy formation models. Dry mergers are considered
as the prime candidates to account for the strong mass and size
evolution of the stellar spheroids at $ z<2$ (van der Wel et al. 2009). As
noted by Bernardi et al. (2007), the
galaxy colour is not expected to change during dry mergers, since
there is no associated star formation, thus galaxies move in the CMR
as the mass of the system increases, but the colour remains fixed.
Skelton et al. (2009) presented a simplified model in which dry
mergers of galaxies already on the red sequence midly change the CMR
slope at higher luminosities, reproducing the change of slope observed
in the bright end of the CMR for the galaxies of the Sloan Digital Sky
Survey.  Their model and results relies on a strong assumption on the
gas fraction threshold chosen to distinguish between dry and wet
mergers. In this article, we use a semi-analytic model of galaxy
formation (Lagos, Cora \& Padilla 2008) to investigate the change of
slope in the bright end of the CMR of early-type cluster galaxies,
taking into account the mass of stars and metals contributed to each
galaxy by quiescent star formation and starburst during mergers and
disk instability events.

\subsection{The Red Sequence}

Our model reproduces very well the general trend of the CMR of cluster
galaxies, as becomes evident from the comparison of the results
obtained from a simulated cluster with virial mass $\approx 1.3 \times
10^{15} \, h^{-1} \, {\rm M}_{\odot}$, and early-type galaxies
observed in the central region of the Antlia cluster (Jim\'enez et
al. 2008). We find that more massive galaxies ($ -23 \leq M_{\rm
T_{\rm 1}} \leq -19 $, in the Washington photometric system) depart
from the linear fit to observed data, displaying an almost constant
colour ($C-T_{\rm 1}\approx 1.7)$, as detected in other clusters.  We
divided the simulated CMR in bins of one magnitude from $M_{\rm T_{\rm
1}} =-16$ to $M_{\rm T_{\rm 1}} =-23$, in order to investigate the
different processes that explain the bluer colours of the brigther
galaxies in our simulation. We analyse the evolution of mass and
metallicity of the stellar component contributed by different
processes: quiescent star formation and starbursts during major wet,
major dry, minor wet and minor dry mergers, and disk instability
events.  Mergers are classified as minor or major according to the
ratio between the baryonic mass of the galaxies involved; if this
ratio is less than $0.3$, the merger is classified as a minor
one. These mergers are then distinguished between dry or wet depending
on the amount of cold gas available in the remnant galaxy; if it is
less than $60$ \% of the baryonic mass, the merger is considered
dry. During dry mergers there is no star formation.

We compute the stellar mass fraction contributed by the different
processes at all magnitude bins, without taking into account quiescent
star formation, which is the process that provides the major
contribution at all magnitude bins at all redshifts. The evolution
with redshift of these fractions is shown in Figure $1$.  We can see
that, for a set of very luminous galaxies ($ -22 \leq M_{\rm T_{\rm
1}} \leq -21$), the most important process that contribute to the
stellar mass at low redshifts is dry merger. Minor wet mergers events
and disk instability are also important but in lower degree.  As we
move in the CMR to lower magnitude bins, the relative importance of
these processes changes. For the least luminous galaxies ($ -17 \leq
M_{\rm T_{\rm 1}} \leq -16$), we find that the most relevant process
for $z \la 4$ is disk instability, followed by minor wet mergers,
minor dry mergers and major dry mergers. This last process is the
least relevant one for all magnitude bins.

\begin{figure}
\begin{center}
\includegraphics[width=0.45\textwidth]{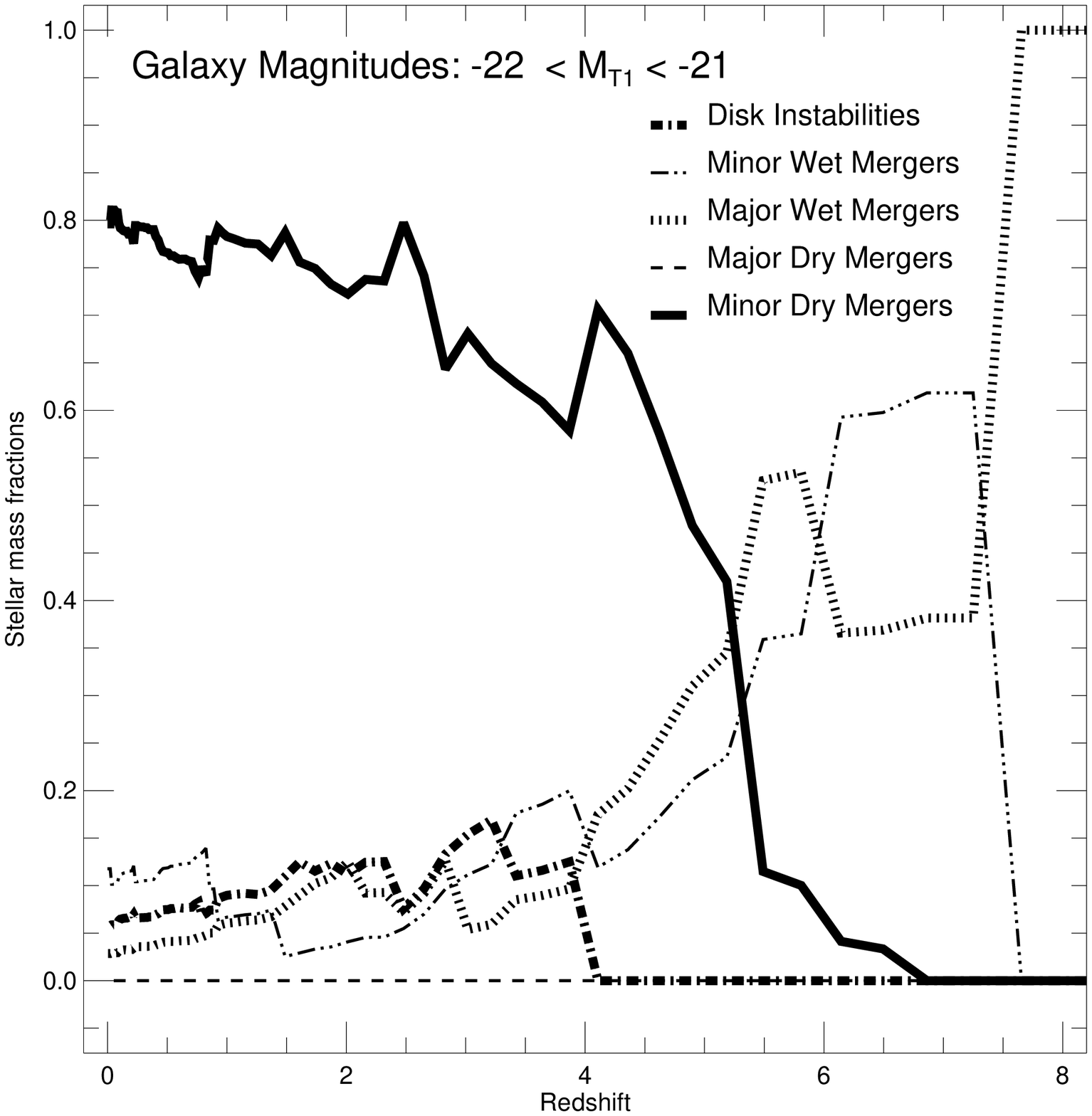}
\includegraphics[width=0.45\textwidth]{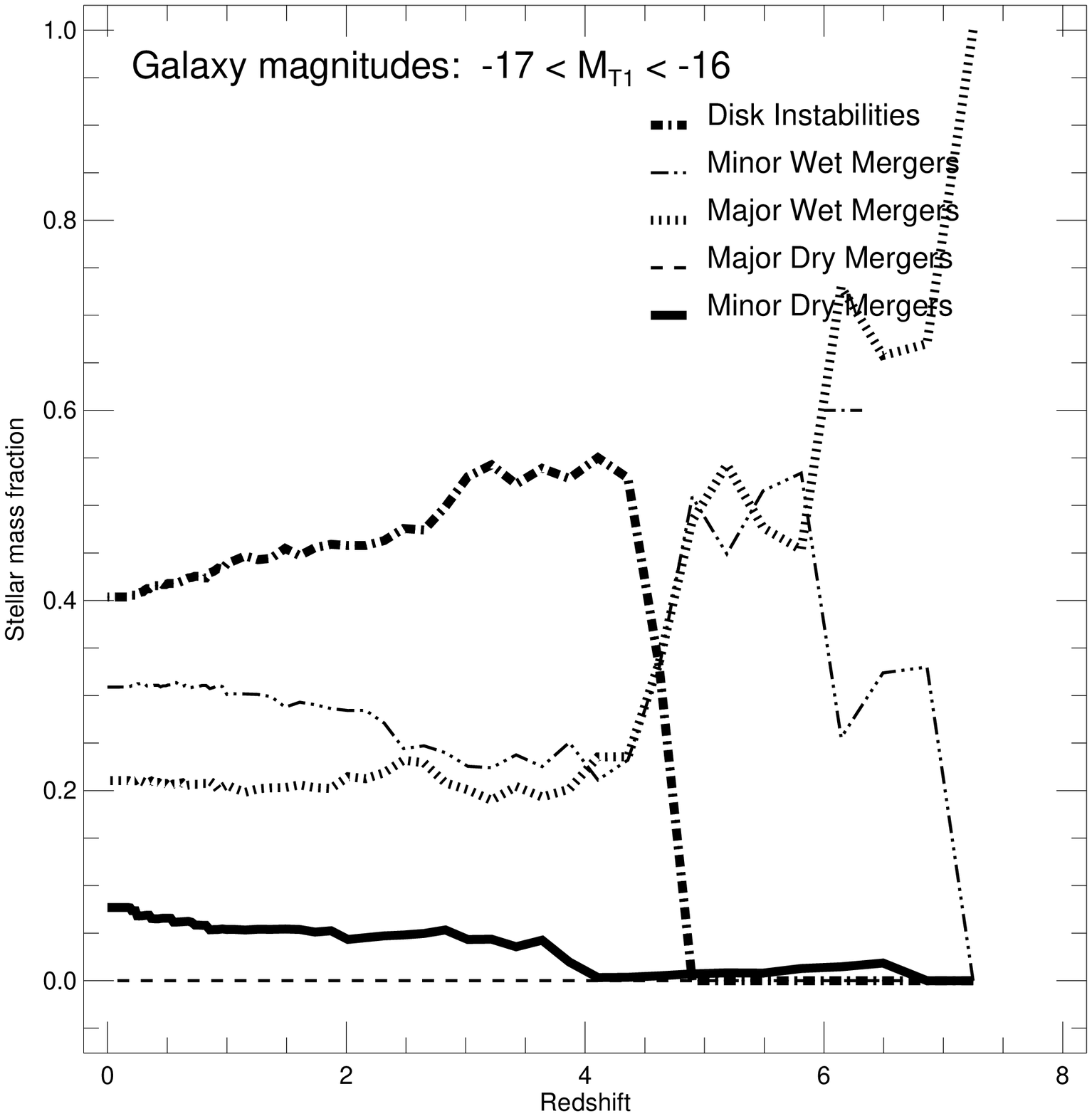}
\caption{{\em Left panel}: Stellar mass fractions due to the different 
processes that lead to galaxy formation 
for one of the first bins of magnitude in the CMR 
($ -22 \leq{\rm T_{\rm 1}} \leq -21 $).
{\em Rigth panels}: Idem as left panel but for the lower magnitude bin
($ -17 \leq  {\rm T_{\rm 1}} \leq -16 $).}
\label{merTotDM}
\end{center}
\end{figure}

\begin{figure}
\begin{center}
\includegraphics[width=0.45\textwidth]{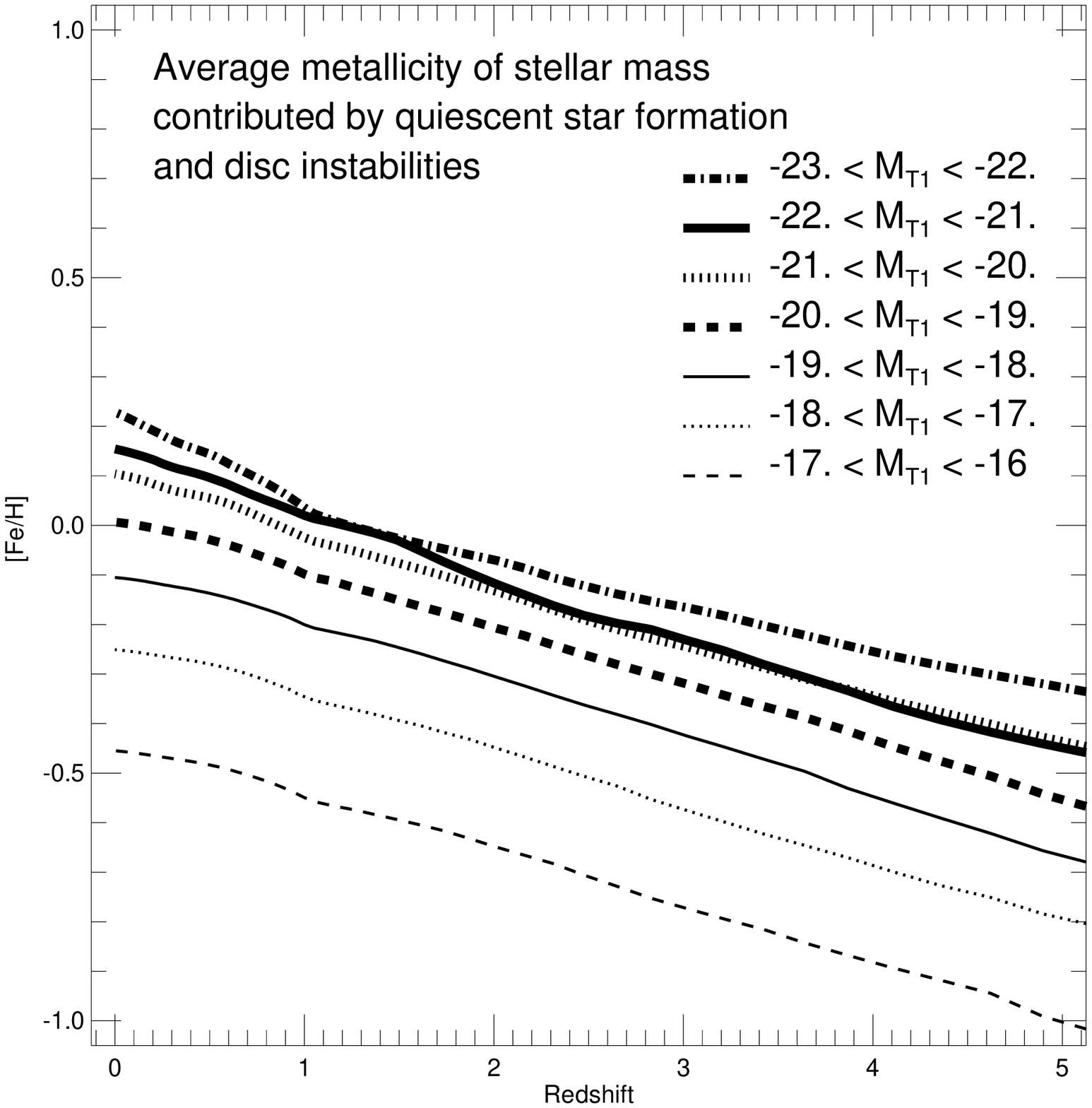}
\includegraphics[width=0.45\textwidth]{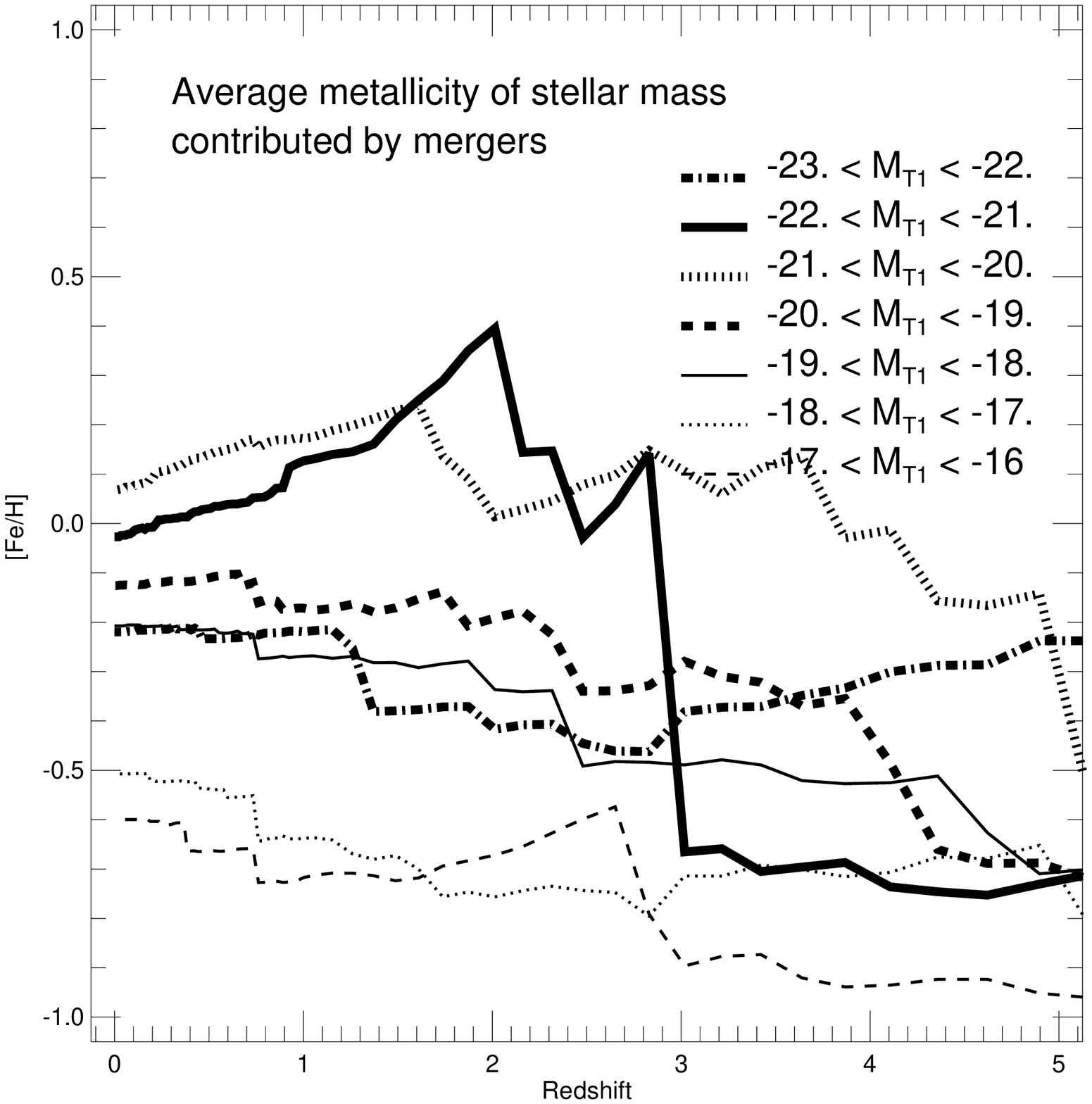}
\caption{Average metallicity for each magnitude bin as a result
of star formation during different processes. 
{\em Left panel}: 
Quiescent star formation and 
disk instability.
{\em Right panel}: All types of  
merger events. 
}
\label{merTotDM}
\end{center}
\end{figure}

We also explore the metallicity of the stars accreted during different
events and estimate the average at each magnitude bin (Figure $2$.)
The average is estimated for two different sets of processes,
quiescent star formation and disk instability, on the one hand, 
and merger event processes, on the other.
Processes in the former set can be considered as internal ones
giving rise to galaxy evolution in isolation, 
while those in the latter are responsible for the external contributions to 
the metal content of the galaxies. 

For galaxies departing from the linear fit ($ -19 \leq M_{\rm T_{\rm
1}} \leq -23$), the average metallicities due to internal processes
increase monotonically with decreasing redshift reaching values
comprised in the range $0 \leq \rm{[Fe/H]} \leq 0.25$ at $z=0$.  We
note that in this case, the more luminous the galaxy, the higher the
achieved metallicity.  Conversely, the stellar mass contributed by
external processes have mainly subsolar metallicities ($ -0.22 \leq
\rm{[Fe/H]}\leq 0.1$), being the most massive galaxies the one
receiving the least amount of metals.  Hence, as a consequence of
mergers, the final metallicity of the more luminous galaxies along the
CMR ($0\leq\rm{[Fe/H]}\leq 0.15 $ at $z=0$), is lower than the one that
would have if only internal processes were acting.  Thus, we find that
the major contribution of stellar mass to the more luminous galaxies
of the CMR at low redshift is due to minor dry mergers (see Figure
$1$), being the metallicity of the accreted stellar mass quite low. It
this way, assuming that colours are mainly driven by metallicity, it
seems that the departure of the bright end of the CMR from a linear
fit can be explained by the effect of minor dry mergers; they would
increase the mass of the galaxies without changing their colours.


\begin{thebibliography}{}
\bibitem[Bower et al. (1992)]{Bower92}
Bower, R. G., Lucey, J. R., \& Ellis, R. S. 1992, MNRAS, 254, 601
\bibitem[Jim\'enez, Cora, Bassino,Smith Castelli (2008)]{Jimenez08}
Jim\'enez, N., Cora S.A., Bassino L.P., Smith Castelli, A. 2008,BAAA, V51,263,266\bibitem[Janz \& Lisker (2009)]{Janz09}
Janz, J. , Lisker, T.  2009,  \apj, 696, 102L
\bibitem[Lagos, Cora \& Padilla (2008)]{Lagos08} 
Lagos, C., Cora S.A., \& Padilla N.D. 2008, \mnras, 388, 587
\bibitem[Skelton et al. (2009)]{Skelton09}
Skelton, R.E., Bell, E.F.,Somerville, R.S. 2009, \apj, 699, 9L
 	
\end{thebibliography}
\end{document}